\documentclass[12pt]{iopart}
\eqnobysec
\usepackage{iopams,amsthm,graphicx,cite}

\begin{document}

%%%%%%%%%%%%---ARROWS FOR CONVERGENCE----%%%%%%%%%%%%%%%%%%%%%%%%%%%%%%%%%%%%%
\def\llra{\relbar\joinrel\longrightarrow}              %THIS IS LONG
\def\mapright#1{\smash{\mathop{\llra}\limits_{#1}}}    %ARROW ON LINE
\def\mapup#1{\smash{\mathop{\llra}\limits^{#1}}}     %CAN PUT SOMETHING OVER IT
\def\mapupdown#1#2{\smash{\mathop{\llra}\limits^{#1}_{#2}}} %over&under it%
%%%%%%%%%%%%%%%%%%%%%%%%%%%%%%%%%%%%%%%%%%%%%%%%%%%%%%%%%%%%%%%%%%%%%%%%%%%%%%

%%%%%%%%%%%%%% These are the AMS constructs for multiline limits %%%%%%%%%%%%
\catcode`\@=11

\def\BF#1{{\bf {#1}}}
\def\NEG#1{{\rlap/#1}}

\def\Let@{\relax\iffalse{\fi\let\\=\cr\iffalse}\fi}
\def\vspace@{\def\vspace##1{\crcr\noalign{\vskip##1\relax}}}
\def\multilimits@{\bgroup\vspace@\Let@
 \baselineskip\fontdimen10 \scriptfont\tw@
 \advance\baselineskip\fontdimen12 \scriptfont\tw@
 \lineskip\thr@@\fontdimen8 \scriptfont\thr@@
 \lineskiplimit\lineskip
 \vbox\bgroup\ialign\bgroup\hfil$\m@th\scriptstyle{##}$\hfil\crcr}
\def\Sb{_\multilimits@}
\def\endSb{\crcr\egroup\egroup\egroup}
\def\Sp{^\multilimits@}
\let\endSp\endSb

%%%%%%%%%%%%%%%%%%%%END of explanations for multiline limits %%%%%%%%%%%%%%%%

\title[]{Reply to Comment~{\sf arXiv:0704.3529v1}}

\author{Rafael de la Madrid}
\address{Department of Physics, University of California at San Diego,
La Jolla, CA 92093 \\
E-mail: \texttt{rafa@physics.ucsd.edu}}

\date{\small{(December 10, 2006)}}

\begin{abstract}
In this reply, we show that the author of the Comment~{\sf arXiv:0704.3529v1}
inadvertently provides additional arguments against the use of Hardy functions 
as test functions for the Gamow states.
\end{abstract}


%\submitto{\JPA}

\section{Introduction}

The author of Comment~\cite{C} asserts that the conclusions of~\cite{HARDY} 
regarding the inconsistency of the ``Hardy axiom'' with quantum mechanics 
are wrong. In
this reply, we will see that the criticisms of~\cite{C} actually provide 
another way to see that Hardy functions should not be used as test functions 
for the Gamow (resonant) states.

In order to show so, we will introduce Hardy functions in two different 
ways. The first way follows the method of~\cite{C}. The second way consists of
applying the method of~\cite{C} to the Lippmann-Schwinger equation, which 
leads to the ``Hardy axiom'' of the Bohm-Gadella (BG) theory. It will then
become apparent that the use of Hardy functions is arbitrary, because we can 
introduce not only Hardy but any type of functions using the method 
of~\cite{C}. Afterward, we will recall~\cite{REPLY} how to construct spaces 
of test functions for the standard Gamow states using standard distribution 
theory. We shall refer to this method as the ``standard method'' and explain 
why it is incompatible with choosing Hardy functions. We shall 
finally point out that the method of~\cite{C} doesn't even lead to analytic 
(let alone Hardy) functions when applied to the test functions
obtained by the ``standard method.'' In the progress of the reply, we will 
overcome all the objections of~\cite{C}.

\section{Two ways to introduce Hardy functions}

For the sake of clarity, we shall use the spherical shell potential of height 
$V_0$,
\begin{equation}
	V(\vec{x})=V(r)=\left\{ \begin{array}{ll}
                                0   &0<r<a  \\
                                V_0 &a<r<b  \\
                                0   &b<r<\infty \, ,
                  \end{array} 
                 \right. 
	\label{potential}
\end{equation}
and restrict ourselves to the zero angular momentum case, although our 
conclusions will be valid for any partial wave and for any smooth potential 
that falls off at infinity faster than any exponential. In the radial position 
representation, the Hamiltonian acts as
\begin{equation}
       H= -\frac{\hbar ^2}{2m} \frac{\rmd ^2}{\rmd r^2}+V(r) \, .
      \label{doh}
\end{equation}
The regular solution is
\begin{equation}
      \chi (r;E)=\left\{ \begin{array}{lll}
               \sin (kr) \quad &0<r<a  \\
               {\cal J}_1(k)\rme ^{\rmi Qr}
                +{\cal J}_2(k)\rme ^{-\rmi Qr}
                 \quad  &a<r<b \\
               {\cal J}_3(k) \rme ^{\rmi kr}
                +{\cal J}_4(k)\rme ^{-\rmi kr}
                 \quad  &b<r<\infty \, ,
               \end{array} 
                 \right. 
             \label{chi}
\end{equation}
where
\begin{equation}
      k=\sqrt{\frac{2m}{\hbar ^2}E \,} \, , \quad 
      Q=\sqrt{\frac{2m}{\hbar ^2}(E-V_0) \,} \, .
\end{equation}
The Jost functions are given by
\begin{equation}
      {\cal J}_+(E)=-2\rmi {\cal J}_4(E) \, , \quad 
      {\cal J}_-(E)=2\rmi {\cal J}_3(E) \, .
        \label{josfuc1}
\end{equation}

It was shown in~\cite{ICTP} that there are (at least) three different,
physically inequivalent eigenfunctions associated with $H$:
\begin{equation}
      \chi _{\rm sw}(r;E)= \sqrt{\varrho _{\rm sw}(E)} \, \chi (r;E) \, ,
      \label{dnesb}
\end{equation}
\begin{equation}
      \chi ^{\pm}(r;E) =\sqrt{\varrho (E)}
                                  \, \frac{\chi (r;E)}{{\cal J}_{\pm}(E)} 
              \, ,  
          \label{LSdnesb}
\end{equation}
where
\begin{equation}
      \varrho _{\rm sw}(E)=
      \frac{1}{\pi}\, 
      \frac{2m/\hbar ^2}{k}
      \, \frac{1}{ |{\cal J}_+(E)|^2}   \, ,
\end{equation} 
\begin{equation}
      \varrho (E) =\frac{1}{\pi}\,\frac{2m/\hbar ^2}{k} \, .
      \label{rph1}
\end{equation} 
The first eigensolution, $\chi _{\rm sw}(r;E)$, was called in~\cite{ICTP} the 
standing-wave eigenfunction, whereas $\chi ^{\pm}(r;E)$ are the 
Lippmann-Schwinger eigenfunctions. All these eigenfunctions are 
delta-normalized
\begin{equation}
    \hskip-0.5cm   \int_0^{\infty}\rmd r \, \overline{\chi _{\rm sw}(r;E')}
                          \chi _{\rm sw}(r;E) =
       \int_0^{\infty}\rmd r \, \overline{\chi ^{\pm}(r;E')}
                          \chi ^{\pm}(r;E)= \delta (E-E') \, .
       \label{detanor1}
\end{equation}
Thus, their associated unitary operators
\begin{equation}
      (U_{\rm sw}f)(E)=\int_0^{\infty}\rmd r \, 
                       \overline{\chi _{\rm sw}(r;E)}\, f(r) 
          \label{Usw}
\end{equation}
\begin{equation}
      (U_{\pm}f)(E)=\int_0^{\infty}\rmd r \, 
                         \overline{\chi ^{\pm}(r;E)}\, f(r) 
         \label{Upm}
\end{equation}
transform from $L^2([0,\infty),\rmd r)$ onto $L^2([0,\infty),\rmd E)$. 

The author of~\cite{C} uses the eigenfunctions $\chi _{\rm sw}(r;E)$ and their
associated unitary operator $U_{\rm sw}$.\footnote{In~\cite{C}, $U_{\rm sw}$
is denoted by $\Psi$.} It is obvious from 
expressions~(\ref{dnesb})-(\ref{Upm}) that working with the 
Lippmann-Schwinger eigenfunctions $\chi ^{\pm}(r;E)$ and $U_{\pm}$ is
mathematically and conceptually as
easy as working with the standing-wave eigenfunctions $\chi _{\rm sw}(r;E)$ 
and $U_{\rm sw}$. What is more, since $\chi _{\rm sw}(r;E)$ does not fulfill 
the scattering boundary conditions (it fulfills ``standing-wave'' boundary
conditions~\cite{ICTP}), it doesn't always lead to physically correct answers 
in scattering theory. Thus, the objection of~\cite{C} to the use of the 
Lippmann-Schwinger equation in scattering theory is unwarranted.

Once we have constructed $U_{\rm sw}$ and $U_{\pm}$, we can apply 
the argument of~\cite{C} to obtain the spaces of test functions for the
Gamow states. We simply have to use $U_{\rm sw}$ and $U_{\pm}$ to transform 
into the energy representation, where we impose the Hardy condition. If we 
denote
by $\tilde{\Phi}_{+}$ ($\tilde{\Phi}_{-}$) the space of Hardy functions from 
above (below) restricted to $[0,\infty )$, then the spaces on which the Gamow 
states supposedly act are\footnote{In the BG theory, one also 
imposes that the elements of $\tilde{\Phi}_{\mp}$ are Schwartz functions.}
\begin{equation}
       \Phi _{{\rm BG}\pm}= U_{\pm}^{-1} \tilde{\Phi}_{\mp}  \, ,  
     \label{BGchoice}
\end{equation}
\begin{equation}
       \Phi _{{\rm H}\pm}= U_{\rm sw}^{-1} \tilde{\Phi}_{\pm} \, .    
     \label{hchoice}
\end{equation}
The {\it choice}~(\ref{BGchoice}) is called the ``Hardy axiom'' in the 
BG theory. The {\it choice}~(\ref{hchoice}) is what the author 
of~\cite{C} calls ``a fact, a theorem.'' It is obvious that the choices
(\ref{BGchoice}) and (\ref{hchoice}) are arbitrary. One may as well
choose another subset of $L^2([0,\infty ),\rmd E)$ with different properties
and obtain a different space of test functions for the Gamow states. What is 
more, $\Phi _{{\rm BG}\pm}$ are different from $\Phi _{{\rm H}\pm}$, and the 
resonant states constructed using the choice of~\cite{C} would, in the position
representation, be different from the resonant states of the BG
theory, which in turn are different from the standard Gamow 
states~\cite{HARDY}.

Even though the author of~\cite{C} asserts that the physical 
consequences of choosing Hardy functions remain within standard quantum 
mechanics, proponents of the BG theory have repeatedly stated that such choice 
goes beyond the framework of traditional quantum mechanics. The main 
differences between standard quantum mechanics and the BG theory 
were presented in~\cite{HARDY}. Thus, it is not only the present author but 
also the proponents of the BG theory who think that using Hardy 
functions has physical consequences that lie beyond standard quantum mechanics.

\section{The shift parameter ``$x$'' vs.~the quantum arrow of time}
\label{sec:xves}

One reason why we would make the choices~(\ref{BGchoice}) or~(\ref{hchoice}) 
is if there was a physical justification for them. In the 
BG theory, the justification for~(\ref{BGchoice}) is provided by the 
``quantum arrow of time.'' For such arrow of time to have a
physical meaning, one has to identify what the author of~\cite{C} calls 
``$x$'' with the time evolution parameter. Such identification is done not 
by~\cite{HARDY}, as the author of~\cite{C} claims, but by~\cite{PLA94,JMP95}, 
where the quantum arrow of time was introduced. What~\cite{HARDY} shows 
is that such identification is not possible and that therefore the 
choice~(\ref{BGchoice}) lacks a physical justification. Thus, the 
criticism of~\cite{C} about the misinterpretation of ``$x$'' by~\cite{HARDY} 
is not only unwarranted, but actually is a recognition that the 
quantum arrow of time of~\cite{PLA94,JMP95} is physically flawed,
in agreement with~\cite{HARDY}.

The author of~\cite{C} also criticizes the following equation:
\begin{equation}
     0=\langle ^+E|\varphi ^{+}(t)\rangle = \rme ^{-\rmi Et} \varphi ^{+}(E)
      \, . 
        \label{flawassum1} 
\end{equation}
The author of~\cite{C} correctly points out that Eq.~(\ref{flawassum1}) 
suggests that ``{\it something is wrong}.'' What the author of~\cite{C} 
doesn't mention is that Eq.~(\ref{flawassum1}) was not proposed 
by~\cite{HARDY} but was used in~\cite{PLA94,JMP95} to establish the quantum 
arrow of time, see~\cite[Eq.~(3.4)]{PLA94} and~\cite[Eq.~(3.4)]{JMP95}.

\section{Analytic and growth properties of the test functions obtained by the
``standard method''}
\label{sec:analid}

Let us now turn to the main conclusion of~\cite{HARDY},
namely that the choice~(\ref{BGchoice}), and as we will see also the
choice~(\ref{hchoice}), is inconsistent with quantum 
mechanics. The ultimate goal for making the assumptions~(\ref{BGchoice}) 
or~(\ref{hchoice}) is to find a space of test functions on which the Gamow 
states act, because mathematically the Gamow states are distributions. Thus,
the ultimate test to check whether the choices~(\ref{BGchoice}) 
or~(\ref{hchoice}) are consistent with quantum mechanics is to show---rather
than to assume---that the standard Gamow states act on the spaces
$\Phi _{{\rm BG}\pm}$ or $\Phi _{{\rm H}\pm}$. With that goal in mind, we
are going to compare the spaces~(\ref{BGchoice}) and~(\ref{hchoice}) with
the spaces of test functions that one finds by applying distribution 
theory~\cite{GELFAND} to the Gamow states. 

The theory of distributions~\cite{GELFAND} says that a test function 
$\varphi (r)$ on which a distribution $d(r)$ acts is such that the following
integral is finite:
\begin{equation}
     \langle \varphi|d\rangle \equiv 
        \int_0^{\infty}\rmd r \, \overline{\varphi (r)} d(r) <\infty \, ,
          \label{basic}
\end{equation}
where $\langle \varphi |d\rangle$ represents the action of the
functional $|d\rangle$ on the test function $\varphi$. With some variations, 
this is the ``standard method'' followed 
by~\cite{SUDARSHAN,BOLLINI,FP02,JPA02,IJTP03,JPA04,EJP05,LS1,LS2} to introduce 
spaces of test functions in quantum mechanics. Since the Gamow eigenfunctions
$u(r;z_n)$ associated with the resonant energies $z_n$ blow up exponentially, 
the basic requirement~(\ref{basic}) of the ``standard method'' tells us that 
the test functions $\varphi (r)$ for the Gamow states must fall off at least 
exponentially. We shall denote the collection of such $\varphi (r)$ by
$\mathbf \Phi$. Using well-known results from distribution 
theory~\cite{GELFAND}, it was 
shown in~\cite{LS2} that when $\varphi$ belongs to $\mathbf \Phi$, the 
analytic continuation of $(U_\pm \varphi )(E)$ tends to infinity in the 
infinite arc of the complex plane, in accordance with an analogous property
of ultradistributions~\cite{BOLLINI}. Thus, the
energy representations of the space $\mathbf \Phi$,
$U_{\pm}{\mathbf \Phi}$, are not characterized by the Hardy-function condition,
\begin{equation}
      U_{\pm}{\mathbf \Phi} \neq \tilde{\Phi}_{\mp} \, , 
            \label{difecd}
\end{equation}
and therefore in the energy representation the standard Gamow states do not 
act on spaces of Hardy functions~\cite{NOTE}. 

The argument leading to~(\ref{difecd}) can also be applied to the 
choice~(\ref{hchoice}) to show that when $\varphi$ belongs to $\mathbf \Phi$, 
the analytic continuation of $(U_{\rm sw} \varphi )(E)$ tends to infinity in 
the infinite arc of the complex plane, thereby showing that the 
spaces~(\ref{hchoice}) are not the same as the spaces of test functions on 
which the standard Gamow states act. We will not need to prove so however, 
because the proposal of~\cite{C} suffers from more serious problems, as 
the next section shows.

\section{Analytic properties of $(U_{\rm sw}f)(E)$}
\label{sec:andifn}

In the BG theory, and in the proposal of~\cite{C}, one works in the energy
(spectral) representation of the Hamiltonian, and therefore one never deals
with the Gamow states introduced originally by Gamow. It is simply assumed
that one can bring the Hardy-function assumption back into the position
representation by way of $U_{\pm}$ or $U_{\rm sw}$, and that everything
works out fine. The purpose of~\cite{HARDY} was to show that such formal 
treatment is simply inconsistent with the ``standard method'' of dealing with 
the Gamow states. What is more, by working that way one may be led to
wrong conclusions. To see why, let us take a closer look at the 
choice~(\ref{hchoice}). We shall write the operator~(\ref{Usw}) in such a
way that we can perform the analytic continuation of $(U_{\rm sw}f)(E)$:
\begin{equation}
      (U_{\rm sw}f)(E)= \left( \frac{1}{\pi}\, 
      \frac{2m/\hbar ^2}{k}
      \, \frac{1}{{\cal J}_+(E){\cal J}_-(E)} \right) ^{1/2} 
         \int_0^{\infty}\rmd r \, \chi (r;E) f(r) \, .
      \label{disection}
\end{equation}
Since the regular solution $\chi (r;E)$ is analytic everywhere for the
kind of potentials we are considering, the integral in Eq.~(\ref{disection})
will yield an analytic function of $E$ when such integral is sufficiently
convergent. The analytic continuation of the factor inside the square root,
\begin{equation}
       \frac{1}{\pi} \frac{2m/\hbar ^2}{k}
      \, \frac{1}{{\cal J}_+(E) \, {\cal J}_-(E)} \, , 
            \label{JFund}     
\end{equation}
has poles in any quadrant of the complex 
$k$-plane~\cite{NUSSENZVEIG,NEWTON}, which implies that the analytic 
continuation of~(\ref{JFund}) has 
poles in all half-planes of the Riemann surface. Thus, the
functions $(U_{\rm sw}f)(E)$ of~(\ref{disection}) cannot be analytic 
in any half-plane of the Riemann surface, which in particular means that they 
cannot be Hardy functions. Hence, the choice~(\ref{hchoice}) is 
inconsistent with the analytic properties of the standard quantum
mechanical functions~(\ref{disection}).

\section{Conclusions}

We have shown that~\cite{C} not only doesn't refute the arguments 
of~\cite{HARDY} but in fact provides another way to see that Hardy functions 
are not appropriate test functions for the standard Gamow 
states. Comment~\cite{C} 
simply provides another arbitrary way to introduce
Hardy (or any other type of) functions in quantum mechanics. Regardless of 
whether one calls it an ``axiom'' or ``a fact, a theorem,'' introducing
Hardy functions by hand is physically inconsistent with standard quantum 
mechanics, albeit mathematically possible. 

We have also shown that~\cite{C} inadvertently acknowledges that 
the quantum arrow of time is physically vacuous, which 
was one of the main conclusions of~\cite{HARDY}. Without such arrow of time, 
there is no physical justification for the BG~theory.


\section*{References}


\begin{thebibliography}{99}


\bibitem{C} H.~Baumgartel, {\sf arXiv:0704.3529v1}.

\bibitem{HARDY} R.~de la Madrid, J.~Phys.~A: Math.~Gen.~{\bf 39}, 9255
(2006); {\sf quant-ph/0606186}.

\bibitem{REPLY} R.~de la Madrid, J.~Phys.~A: Math.~Theo.~{\bf 40}, 4671
(2007); {\sf arXiv:0704.1613}.

\bibitem{ICTP} R.~de la Madrid, \emph{The importance of boundary conditions 
in Quantum Mechanics}, in {\it Irreversible Quantum Dynamics}, F.~Benatti, 
R.~Floreanini [Eds.], p.~327, Springer Lecture Notes in Physics, Springer 
(2003); {\sf quant-ph/0302184}.

\bibitem{PLA94} A.~Bohm, I.~Antoniou, P.~Kielanowski, Phys.~Lett.~A~{\bf 189},
442 (1994).

\bibitem{JMP95} A.~Bohm, I.~Antoniou, P.~Kielanowski, J.~Math.~Phys.~{\bf 36},
2593 (1995).

\bibitem{GELFAND} I.M.~Gelfand, G.E.~Shilov {\it Generalized Functions, 
Vol.~1-3}, Academic Press, New York (1964). I.M.~Gelfand, N.Ya.~Vilenkin 
{\it Generalized Functions Vol.~4}, Academic Press, New York (1964).

\bibitem{SUDARSHAN} G.~Parravicini {\it et al.}, 
J.~Math.~Phys.~{\bf 21}, 2208 (1980).

%\bibitem{SUDARSHAN} G.~Parravicini, V.~Gorini, E.C.G.~Sudarshan, 
%J.~Math.~Phys.~{\bf 21}, 2208 (1980).

\bibitem{BOLLINI} C.G.~Bollini, O.~Civitarese, A.L.~De Paoli, M.C.~Rocca,
Phys.~Lett.~B{\bf 382}, 205 (1996); J.~Math.~Phys.~{\bf 37}, 4235 (1996).

\bibitem{FP02} R.~de la Madrid, A.~Bohm, M.~Gadella, 
Fortschr.~Phys.~{\bf 50}, 185 (2002); {\sf quant-ph/0109154}.

\bibitem{JPA02} R.~de la Madrid, J.~Phys.~A:~Math.~Gen.~{\bf 35}, 319 (2002); 
{\sf quant-ph/0110165}.

\bibitem{IJTP03} R.~de la Madrid, Int.~J.~Theo.~Phys.~{\bf 42}, 2441 (2003);
{\sf quant-ph/0210167}.

\bibitem{JPA04} R.~de la Madrid, J.~Phys.~A: Math.~Gen.~{\bf 37}, 8129 (2004); 
{\sf quant-ph/0407195}.

\bibitem{EJP05} R.~de la Madrid, Eur.~J.~Phys.~{\bf 26}, 287 (2005);
{\sf quant-ph/0502053}.

\bibitem{LS1} R.~de la Madrid, J.~Phys.~A:~Math.~Gen.~{\bf 39}, 3949 (2006);
{\sf quant-ph/0603176}. 

\bibitem{LS2} R.~de la Madrid, J.~Phys.~A:~Math.~Gen.~{\bf 39}, 3981 (2006);
{\sf quant-ph/0603177}. 

\bibitem{NOTE} This shows what~\cite{HARDY} meant by the assertion that 
``{\it the limits (2.18) and (2.19) are in general not 
zero}.'' Such limits are in general not zero for the test functions obtained 
by the ``standard method,'' although it is possible to find 
functions of $L^2([0,\infty ), \rmd E)$ for which those limits are zero. 

\bibitem{NUSSENZVEIG} H.M.~Nussenzveig, {\it Causality and 
Dispersion Relations}, Academic Press, New York and London (1972).

\bibitem{NEWTON} R.G.~Newton, 
{\it Scattering Theory of Waves and Particles}, McGraw-Hill,
New York (1966); 2nd edition, Springer Verlag, New York (1982).




\end{thebibliography}
\end{document}